\documentstyle[12pt]{article}
\begin{document}
\def\bt{\begin{tabular}}
\def\et{\end{tabular}}
\def\bfr{\begin{flushright}}
\def\mm{\mbox{\boldmath $ }}
\def\efr{\end{flushright}}
\def\bfl{\begin{flushleft}}
\def\efl{\end{flushleft}}
\def\vs{\vspace}
\def\hs{\hspace}
\def\sta{\stackrel}
\def\pb{\parbox}
\def\bc{\begin{center}}
\def\ec{\end{center}}
\def\sp{\setlength{\parindent}{2\ccwd}}
\def\bp{\begin{picture}}
\def\ep{\end{picture}}
\def\uni{\unitlength=1mm}
\def\REF#1{\par\hangindent\parindent\indent\llap{#1\enspace}\ignorespaces}

\noindent \bc {\LARGE\bf Revised Iterative Solution for \\
\vspace{.2cm} Groundstate of Schroedinger Equation}

\vs*{5mm} {\large   Zhao Wei-Qin$^{1,~2}$}

\vs*{11mm}

{\small \it 1. China Center of Advanced Science and Technology (CCAST)}

{\small \it (World Lab.), P.O. Box 8730, Beijing 100080, China}

{\small \it 2. Institute of High Energy Physics, Chinese Academy of Sciences,

P. O. Box 918(4-1), Beijing 100039, China}

\ec
\vspace{2cm}
\begin{abstract}

A revised iterative method based on Green function defined by
quadratures along a single trajectory is proposed to solve the
low-lying quantum wave function for Schroedinger equation.
Specially a new expression of the perturbed energy is obtained,
which is much simpler than the traditional one. The method is
applied to solve the unharmonic oscillator potential. The revised
iteration procedure gives exactly the same result as those based
on the single trajectory quadrature method. A comparison of the
revised iteration method to the old one is made using the example
of Stark effect. The obtained results are consistent to each other
after making power expansion.
\end{abstract}
\vspace{.5cm}

PACS{:~~03.65.Ge,~~02.30.Mv}

Key words: single trajectory quadrature, Green function, revised
iterative solution

\newpage

\section*{\bf 1. Introduction}
\setcounter{section}{1}
\setcounter{equation}{0}

Recently a single trajectory quadrature method is proposed to
solve the low-lying quantum states of N-dimensional Schroedinger
equation[1,2]. The ground state wave function in N-dimension can
be expressed by quadratures along the single trajectory.
Furthermore a Green function is defined along the single
trajectory[2]. This makes it possible to develop an iterative
method for obtaining the ground state wave function, starting from
a properly chosen trial function. The method is applied to solve
one dimensional double-well potential[3]. The convergence of the
iterative solution very much depends on the choice of the trial
function.

However, the solution based on this iterative formula is not
completely the same as the one obtained from the single trajectory
quadrature method. In the single trajectory quadrature method the
wave function for the ground state is proposed to be in the form
of an exponential, as it should be, while in the original
iterative solution the obtained correction to the wave function is
in the form of a power expansion. This sometimes leads to
unreasonable results, as will be shown in the following.

In this work a revised version of the iterative solution is
introduced, which is based on a different integral equation using
the same set of Green functions. A new  and much simpler
expression of the energy correction is obtained. The solution is
in an exponential form and the result is exactly the same as the
one obtained based on the single trajectory method.

In Section 2, a brief introduction is given about the Green
function method based on the single trajectory quadrature. Special
discussion is given for the revision of the iterative formula. A
comparison of the original and the revised version of the
iterative solution is shown for a simple example. The revised
iterative formula is then applied to the unharmonic oscillator
potential and a comparison of the two iteration solutions is made
using the example of the stark effect in Section 3. Finally some
discussions are given at the end.

\newpage

\section*{\bf 2. Green Function and the Revised Iterative
Solution}
\setcounter{section}{2}
\setcounter{equation}{0}

\vspace{.5cm}

\noindent
{\bf (1). Green function and the revised iterative formula}

We discuss a particle with unit mass, moving in an N-dimensional
unperturbed potential $V_0({\bf q})$. The ground state wave
function $\Phi({\bf q})$ satisfies the following Schroedinger
equation:
\begin{eqnarray}\label{e2.1}
H\Phi({\bf q}) = E \Phi({\bf q}),
\end{eqnarray}
where
\begin{eqnarray}\label{e2.2}
H &=& T+V_0({\bf q})= -\frac{1}{2} {\bf \nabla}^2 + V_0({\bf q}).
\end{eqnarray}
Assume the solution of Eq.(\ref{e2.1}) could be expressed as
\begin{eqnarray}\label{e2.3}
\Phi({\bf q}) &=& e^{-S({\bf q})},
\end{eqnarray}
where $S({\bf q})$ is the trajectory based on which the Green
function will be defined. From Eqs.(\ref{e2.1})-(\ref{e2.3}) it is
easy to derive the following equation for $S({\bf q})$:
\begin{eqnarray}\label{e2.4}
\frac{1}{2}({\bf \nabla}S)^2 - \frac{1}{2}{\bf \nabla}^2S -V_0 + E
= 0.
\end{eqnarray}

Introduce perturbed potential $U({\bf q})$ and assume
\begin{eqnarray}\label{e2.5}
V({\bf q})&=&V_0({\bf q})+U({\bf q})\\ {\cal H} &=& T+V({\bf q})=
-\frac{1}{2} {\bf \nabla}^2 + V({\bf q}).
\end{eqnarray}
Define another wave function $\Psi({\bf q})$ satisfying the
Schroedinger equation
\begin{eqnarray}\label{e2.7}
{\cal H}\Psi({\bf q}) &=& {\cal E} \Psi({\bf q})\\
{\cal E} &=& E+\Delta.
\end{eqnarray}
Let
\begin{eqnarray}\label{e2.9}
\Psi({\bf q}) &=& e^{-S({\bf q})-\tau ({\bf q})}.
\end{eqnarray}
The equation for $\tau$
and $\Delta$ could be derived easily[2]:
\begin{eqnarray}\label{e2.10}
{\bf \nabla}S\cdot {\bf \nabla}\tau + \frac{1}{2}[({\bf
\nabla}\tau)^2 - {\bf \nabla}^2 \tau] = (U - \Delta).
\end{eqnarray}
Consider the coordinate transformation
\begin{eqnarray}\label{e2.11}
q_1, q_2, q_3, \cdots, q_N \rightarrow S, \alpha_1, \alpha_2,
\cdots, \alpha_{N-1}
\end{eqnarray}
with $\alpha=(\alpha_1, \alpha_2, \cdots, \alpha_{N-1})$ denoting the set
of $N-1$ orthogonal angular coordinates satisfying the condition
\begin{eqnarray}\label{e2.12}
{\bf \nabla} S\cdot {\bf \nabla} \alpha_i = 0,
\end{eqnarray}
for $i=1,~2,~\cdots~,N-1$.
Similarly to the discussions in Ref.[2],
introducing the $\theta$-function in S-space:
\begin{eqnarray}\label{e2.13}
\theta(S- {\overline S})= \left\{\begin{array}{cc}
1 & ~~~~~~~~{\sf if}\hspace{4mm} 0 \leq {\overline S} < S \\
0 & ~~~~~~~~{\sf if} \hspace{4mm} 0 \leq S < {\overline S}
\end{array}
\right.
\end{eqnarray}
and define
\begin{eqnarray}\label{e2.14}
C = \theta [({\bf \nabla}S)^2]^{-1}.
\end{eqnarray}
Using
\begin{eqnarray}\label{e2.15}
{\bf \nabla} S \cdot {\bf \nabla} C = 1,
\end{eqnarray}
it is easy to derive the following equation
\begin{eqnarray}\label{e2.16}
\tau = \overline{G} [(U-\Delta)-\frac{1}{2}({\bf
\nabla}\tau)^2]=(1+CT)^{-1}C[(U-\Delta)-\frac{1}{2}({\bf
\nabla}\tau)^2],
\end{eqnarray}
where the Green function $\overline{G}=(1+CT)^{-1} C$.
From this equation and the expression of
\begin{eqnarray}\label{e2.17}
\Delta = \frac{\int \Phi^2({\bf q})U({\bf q})e^{-\tau({\bf
q})}d{\bf q}} {\int \Phi^2({\bf q})e^{-\tau({\bf q})}d{\bf q}}~,
\end{eqnarray}
the iteration process could be performed in
the following way: Starting from $\Delta_0 = 0$ and $\tau_0 = 0$,
\begin{eqnarray}\label{e2.18}
\Delta_n &=& \frac{\int \Phi^2({\bf q}) U e^{-\tau_{n-1}}}
{\int \Phi^2({\bf q}) e^{-\tau_{n-1}} d {\bf q}},\\
\tau_n &=& (1+CT)^{-1} C [(U-\Delta_n)-\frac{1}{2}
({\bf \nabla} \tau_{n-1})^2].
\end{eqnarray}
When the N-dimensional variable ${\bf q}$ is transformed into
$(S,~\alpha)$, $T=-\frac{1}{2}{\bf \nabla}^2$ could be decomposed
into two parts:
\begin{eqnarray}\label{e2.20}
T = T_S + T_{\alpha},
\end{eqnarray}
where $ T_S$ and $T_{\alpha}$ consist only the differentiation to
$S$ and $\alpha$, respectively. The detailed expression of $ T_S$
and $T_{\alpha}$ could be found in Appendix A[2]. Now another
function could be defined as[2]
\begin{eqnarray}\label{e2.21}
\overline{D} &\equiv& -2\theta e^{2 S} \frac{h_S}{h_{\alpha}}
\theta e^{-2 S} h_Sh_{\alpha}
\end{eqnarray}
and it is related to the Green function $\overline{G}$ and $C$ in
the following way[2]:
\begin{eqnarray}\label{e2.22}
\overline{G}=(1+ \overline{D}T_{\alpha})^{-1}\overline{D}=(1+CT)^{-1}C.
\end{eqnarray}
Thus, the integral equation for $\tau$ and $\Delta$ could also be
expressed as
\begin{eqnarray}\label{e2.23}
\tau =\overline{G}[(U-\Delta)-\frac{1}{2}({\bf \nabla}\tau)^2]&=&
(1+\overline{D}T_\alpha)^{-1} \overline{D}
[(U-\Delta)-\frac{1}{2}({\bf \nabla}\tau)^2]\nonumber\\
     &=& \overline{D}(1+T_\alpha \overline{D})^{-1}
[(U-\Delta)-\frac{1}{2}({\bf \nabla}\tau)^2].
\end{eqnarray}
The explicit expression of $\tau$ is
\begin{eqnarray}\label{e2.24}
\tau =-2\int_{0}^{S}e^{2S'} \frac{h_{S' }}{h_{\alpha}}dS'
\int_{0}^{S'}e^{-2S''} h_{S''} h_{\alpha}dS''(1+T_\alpha
\overline{D})^{-1} [(U-\Delta)-\frac{1}{2}({\bf \nabla}\tau)^2].
\end{eqnarray}
Therefore, we have
\begin{eqnarray}\label{e2.25}
-\frac{1}{2} \frac{h_{\alpha} }{h_{S} } e^{-2S} \frac{\partial
\tau(S,~\alpha)}{\partial S} =\int_{0}^{S}e^{-2S'} h_{S'}
h_{\alpha}dS'(1+T_\alpha \overline{D})^{-1}
[(U-\Delta)-\frac{1}{2}({\bf \nabla}\tau)^2].
\end{eqnarray}
The left hand side of Eq.(\ref{e2.25})   approaches $0$ when $S
\rightarrow \infty$, so is the right hand side, i.e.,
\begin{eqnarray}\label{e2.26}
\int_{0}^\infty e^{-2S} h_{S} h_{\alpha}dS(1+T_\alpha
\overline{D})^{-1} [(U-\Delta)-\frac{1}{2}({\bf \nabla}\tau)^2]=0,
\end{eqnarray}
which is correct  for all $\alpha$. Integrating over
$d\alpha=\Pi_{i=1}^{N-1}d\alpha_i$ and because of
\begin{eqnarray}\label{e2.27}
 h_{S} h_{\alpha}T_\alpha=-\frac{1}{2}
\sum_{j=1}^{N-1} \frac{\partial}{\partial \alpha_j} \frac{h_{S}
h_{\alpha}}{h_j^2}\frac{\partial}{\partial \alpha_j}
\end{eqnarray}
and
\begin{eqnarray}\label{e2.28}
\int d \alpha h_{S} h_{\alpha}T_\alpha \tau=0,
\end{eqnarray}
we derive
\begin{eqnarray}\label{e2.29}
\int h_{S} h_{\alpha}d\alpha dS e^{-2S}
[(U-\Delta)-\frac{1}{2}({\bf \nabla}\tau)^2]=0.
\end{eqnarray}
Denoting $d {\bf q}=h_{S} h_{\alpha}d\alpha dS $ we reach a new
expression of the perturbative energy
\begin{eqnarray}\label{e2.30}
\Delta&=&\frac{\int d{\bf q}~e^{-2S} [U-\frac{1}{2}({\bf
\nabla}\tau)^2]}{\int d{\bf q}~e^{-2S}}\nonumber\\
      &=& \frac{\int d{\bf q}~\Phi^2 [U-\frac{1}{2}({\bf
\nabla}\tau)^2]}{\int d{\bf q}~\Phi^2}~.
\end{eqnarray}
Based on Eqs.(\ref{e2.23}) and (\ref{e2.30}) we have another
iteration series
\begin{eqnarray}\label{e2.31}
\Delta_n&=&\frac{\int d{\bf q}~\Phi^2 [U-\frac{1}{2}({\bf
\nabla}\tau_{n-1})^2]}{\int d{\bf q}~\Phi^2}~,\\
\tau_n &=& (1+\overline{D}T_\alpha)^{-1} \overline{D}
[(U-\Delta_n)-\frac{1}{2}({\bf \nabla}\tau_{n-1})^2].
\end{eqnarray}
Or, we can also use Eqs.(\ref{e2.31}) and (2.19), instead of
Eqs.(\ref{e2.18}) and (2.19), as our iteration series, since the
calculation of Eq.(\ref{e2.31}) is much simpler than
Eq.(\ref{e2.18}).

 Comparing the
above iteration with the old one derived from the equations for
$f=e^{-\tau}$ and $\Delta$[2]:
\begin{eqnarray}\label{e2.33}
f_n &=& 1+\overline{G}(-U+\Delta_n)f_{n-1}~,\nonumber\\
\Delta_n&=&\frac{\int d{\bf q}~\Phi^2 Uf_{n-1}}{\int d{\bf
q}~\Phi^2 f_{n-1}}~,
\end{eqnarray}
there are several advantages for the new one:

1) It directly gives an exponential form for the perturbed wave
function $e^{-\tau}$. This result is consistent with those
obtained in the single trajectory quadrature method using the
series expansion of $\{S_i\}$ and $ \{E_i\}$ (See Section 1 of
Ref.[2]).

2) It makes the iteration more transparent, since there are no
terms appearing as $U f$ in the integration.

3) The calculation of the perturbation energy is much simpler.

\vspace{.5cm}

\noindent
{\bf (2). Example}

This formula has been tried for several examples. In the following
the result for a simple example will be given, with the
one-dimensional potential
\begin{eqnarray}\label{e2.34}
V(x)=\frac{1}{2}g^2x^2 + \lambda x~.
\end{eqnarray}
Now we have
\begin{eqnarray}\label{e2.35}
V_0(x)&=& \frac{1}{2}g^2x^2,\\
U(x) &=&  \lambda x~
\end{eqnarray}
and according to Eqs.(\ref{e2.1}) and (\ref{e2.3})
\begin{eqnarray}\label{e2.37}
\Phi(x) &=& e^{-\frac{1}{2}g x^2},\\
S(x)&=& \frac{1}{2}g x^2.
\end{eqnarray}
From Eqs.(\ref{e2.14}) and (\ref{e2.15}) we have, in this example
\begin{eqnarray}\label{e2.39}
C &=& \theta (\frac{\partial S}{\partial x})^{-2}\nonumber\\
  &=& \int\limits_0^S dS~\frac{1}{(gx)^2}\nonumber\\
  &=& \int\limits_0^x \frac{1}{gx}~dx~.
\end{eqnarray}
Define
\begin{eqnarray}\label{e2.40}
\Psi(x) = e^{-\frac{1}{2} g x^2-\tau (x)}.
\end{eqnarray}
Starting from $\tau_0=0$ and $\Delta_0=0$, based on
Eqs.(\ref{e2.18}) and (2.19), we have
\begin{eqnarray}\label{e2.41}
\Delta_1&=&\frac{\int\limits_{-\infty}^{\infty}e^{-gx^2}\lambda x dx}
{\int\limits_{-\infty}^{\infty}e^{-g x^2} dx}~=~0
\end{eqnarray}
and
\begin{eqnarray}\label{e2.42}
\tau_1 &=& C U = \frac{\lambda}{g} x.
\end{eqnarray}
The next step gives
\begin{eqnarray}\label{e2.43}
\Delta_2&=&\frac{\int\limits_{-\infty}^{\infty}e^{-gx^2}\lambda x e^{-\frac{\lambda}{g}x} dx}
{\int\limits_{-\infty}^{\infty}e^{-g x^2}e^{-\frac{\lambda}{g}x} dx}~\nonumber\\
&=&-\frac{\lambda^2}{2 g^2}.
\end{eqnarray}
While further iteration gives
\begin{eqnarray}\label{e2.44}
\tau_2 &=& (1+CT)^{-1} C [(U-\Delta_2)-\frac{1}{2}
({\bf \nabla} \tau_1)^2] = \tau_1.
\end{eqnarray}
The iteration is completed here and the obtained corrections for the
ground state energy and wave function are
\begin{eqnarray}\label{e2.45}
\Delta &=& -\frac{\lambda^2}{2 g^2},\\
\tau &=& \frac{\lambda}{g} x.
\end{eqnarray}
This is the exact solution of the ground state for the potential
of Eq.(\ref{e2.34}).

Performing the iteration based on Eqs.(\ref{e2.31}) and (2.32),
starting again from $\tau_0=0$ and $\Delta_0=0$, we have
\begin{eqnarray}\label{e2.47}
\overline{D}=-2 \int\limits_0^x e^{gy^2} dy
\int\limits_{-\infty}^y e^{-gz^2}dz.
\end{eqnarray}
$\Delta_1=0$ and
\begin{eqnarray}\label{e2.48}
\tau_1 &=& \overline{D}U = \frac{\lambda}{g} x.
\end{eqnarray}
The expression of $\Delta_2$ is
\begin{eqnarray}\label{e2.49}
\Delta_2&=&\frac{\int\limits_{-\infty}^{\infty}e^{-gx^2}(\lambda x
-\frac{1}{2} \frac{\lambda^2}{g^2}) dx}
{\int\limits_{-\infty}^{\infty}e^{-g x^2} dx}~\nonumber\\
&=&-\frac{\lambda^2}{2 g^2},
\end{eqnarray}
which is the same as the result in Eq.(\ref{e2.43}) and
\begin{eqnarray}\label{e2.50}
\tau_2 &=& \overline{D}[(U-\Delta_2)-\frac{1}{2}({\bf \nabla}
\tau_1)^2]\nonumber\\
    &=& \overline{D} U = \tau_1.
\end{eqnarray}
The iteration is completed and the exact solution is obtained
after only two steps of iteration. There is no restriction of the
parameter $\lambda$ to obtain this result.

If applying the old iteration based on Eq.(\ref{e2.33}) for $f_n$
and $\Delta_n$, the obtained corrections for the ground state
energy would be expressed as a ratio of two polynomials, while the
wave function would be a polynomial. Although one seems to reach
the correct energy correction of $\Delta_2=-\frac{\lambda^2}{2
g^2}$ in the second order of the iteration, the equation similar
to Eq.(2.50) never appears between $f_2$ and $f_1$, which means
that the iteration would never complete. The $f_n$ becomes more
and more complicated with larger $n$, so is $\Delta_n$. Only when
$\lambda^2<< g^3$, these expressions are meaningful. Under this
condition, after making the series expansion, the expression could
approach the series expansion of the exact solution. Therefore,
the simple expression of the exact solution could not be reached
in the old iteration procedure.

\newpage

\section*{\bf 3. Application of the Revised Iterative Formula }
\setcounter{section}{3}
\setcounter{equation}{0}

In this section the revised iterative formula is applied to solve
the anharmonic oscillator potential. The two iteration procedures
are compared in details using the Stark effect as an example.\\

\noindent
{\bf 3.1 Anharmonic Oscillator potential}\\

As an example we discuss the one-dimensional harmonic oscillator
with perturbative potentials, applying the revised iterative
formula based on the Green functions defined in Eqs.(2.14), (2.21)
and (2.22). And the the result is compared to those obtained in
ref.[2] using the
perturbation expansion based on the same set of Green functions.\\

\noindent
\underline{Example 1}.~~~~~Consider a one-dimensional harmonic oscillator
with
\begin{eqnarray}\label{e3.1}
H = -\frac{1}{2} \frac{d^2}{dx^2} + \frac{1}{2}~g^2 x^2
\end{eqnarray}
and a perturbative potential
\begin{eqnarray}\label{e3.2}
 U = \epsilon~x^{2p},
\end{eqnarray}
where $p$ is a positive integer. The unperturbed ground-state
wave function is
\begin{eqnarray}\label{e3.3}
\Phi = e^{- S} = e^{- \frac{1}{2}~gx^2 },
\end{eqnarray}
which satisfies Schroedinger equation
\begin{eqnarray}\label{e3.4}
H\Phi(x)=E\Phi(x).
\end{eqnarray}
Since $ S =  \frac{1}{2}~g x^2$, we have
\begin{eqnarray}\label{e3.5}
h_S^2 = (dS/dx)^{-2} = (gx)^{-2} = (2gS)^{-1}.
\end{eqnarray}
The eigenstate of
\begin{eqnarray}
{\cal H} = -\frac{1}{2}\frac{d^2}{dx^2} + \frac{1}{2}~g^2 x^2 + \epsilon~x^{2p}
\nonumber
\end{eqnarray}
is $\Psi(x)= e^{-S-\tau}$ which satisfies Schroedinger equation
\begin{eqnarray}\label{e3.6}
{\cal H}\Psi(x)=(E+\Delta)\Psi(x).
\end{eqnarray}

For one-dimensional case, Eq.(2.22) becomes
\begin{eqnarray}\label{e3.7}
\overline{G}=\overline{D}=(1+CT)^{-1}C.
\end{eqnarray}
The revised iterative formula now consist of Eqs.(2.19) or (2.32)
for $\tau_n$ and Eq.(2.31) for $\Delta_n$ as following:
\begin{eqnarray}\label{e3.8}
\Delta_n&=&\frac{\int\limits_{-\infty}^{\infty} dx~e^{-gx^2} [~\epsilon~x^{2p}-
\frac{1}{2}~(\tau'_{n-1})^2~]}{\int\limits_{-\infty}^{\infty} dx~e^{-gx^2}}~,\\
\tau_n &=& (1+CT)^{-1}C~[~\epsilon~x^{2p}-\Delta_n-
\frac{1}{2}~(\tau'_{n-1})^2~]\nonumber \\
       &=& \overline{D}~[~\epsilon~x^{2p}-\Delta_n-
\frac{1}{2}~(\tau'_{n-1})^2~].
\end{eqnarray}
Taking
\begin{eqnarray}\label{e3.10}
\tau_0(x)=0,~~~~~~~\Delta_0=0,
\end{eqnarray}
based on the formula given in Appendix B, it is easy to obtain
\begin{eqnarray}\label{e3.11}
\Delta_1&=&\frac{\int\limits_{-\infty}^{\infty} dx~e^{-gx^2} \epsilon~x^{2p}
}{\int\limits_{-\infty}^{\infty} dx~e^{-gx^2}}=\epsilon \Gamma_{1~p}~,\\
\tau_1 &=& (1+CT)^{-1}C~[~\epsilon~x^{2p}-\epsilon \Gamma_{1~p}~]\nonumber \\
       &=& \overline{D}~[~\epsilon~x^{2p}-\epsilon \Gamma_{1~p}~]\nonumber \\
       &=& \epsilon \sum_{m=1}^p \Gamma_{mp}~x^{2m}
\end{eqnarray}
and
\begin{eqnarray}\label{e3.13}
\tau'_1 = \epsilon \sum_{m=1}^p 2m \Gamma_{mp}~x^{2m-1}.
\end{eqnarray}
For the second order, we have
\begin{eqnarray}\label{e3.14}
\Delta_2&=&\frac{\int\limits_{-\infty}^{\infty} dx~e^{-gx^2}~[~\epsilon~x^{2p}-
\frac{1}{2}~(\tau'_{1})^2~]}{\int\limits_{-\infty}^{\infty} dx~e^{-gx^2}}\nonumber \\
        &=&\epsilon \Gamma_{1~p}-
        \frac{1}{2}~\epsilon^2 \sum_{m=1}^p\sum_{n=1}^p 4mn\Gamma_{mp}\Gamma_{np}\Gamma_{1,m+n-1}.
\end{eqnarray}
The iteration could continue with $\tau_2$, $\Delta_3$, etc. Finally this will lead to
the energy correction $\Delta_n$ and the wave function in
the order of $n$:
\begin{eqnarray}\label{e3.15}
\Psi_n(x)=e^{-S-\tau_n}.
\end{eqnarray}

Specially, for $p=2$
\begin{eqnarray}
\Delta_2=\epsilon~\frac{3}{(2g)^2}-
         \epsilon^2~\frac{21}{8g^5}.\nonumber
\end{eqnarray}
The result is the same as those from the single trajectory quadrature
method introduced earlier in ref.[2]. When expanding $e^{-\tau}$
to a power series, it is consistent to the perturbation expansion
results obtained in ref.[2] based on the same set of Green function.
\\

\noindent
\underline{Example~2}~~~~Using the same one-dimensional
harmonic oscillator Hamiltonian (3.1) as the unperturbed
$H$, we consider now an odd power perturbation potential
\begin{eqnarray}\label{e3.16}
\epsilon U = \epsilon~x^{2p+1}.
\end{eqnarray}
With analysis parallel to that in the first example,
using the formula introduced in Appendix B,
starting from
\begin{eqnarray}\label{e3.17}
\tau_0(x)=0,~~~~~~~\Delta_0=0,
\end{eqnarray}
it is easy to obtain
\begin{eqnarray}\label{e3.18}
\Delta_1&=&0~,\\
\tau_1 &=& (1+CT)^{-1}C~[~\epsilon~x^{2p+1}~]\nonumber \\
       &=& \overline{D}~[~\epsilon~x^{2p+1}~]\nonumber \\
       &=& \epsilon \sum_{m=0}^p \gamma_{mp}~x^{2m+1}
\end{eqnarray}
and
\begin{eqnarray}\label{e3.20}
\tau'_1 =\epsilon \sum_{m=0}^p (2m+1)\gamma_{mp}~x^{2m}.
\end{eqnarray}
Similarly to Eq.(3.14) we have, for $p\neq 0$,
\begin{eqnarray}\label{e3.21}
\Delta_2&=&\frac{\int dx~e^{-gx^2}~[~\epsilon~x^{2p+1}-
\frac{1}{2}~(\tau'_{1})^2~]}{\int dx~e^{-gx^2}}\nonumber \\
        &=&-\frac{1}{2}~\epsilon^2 \sum_{m=0}^p\sum_{n=0}^p
        (2m+1)(2n+1)\gamma_{mp}\gamma_{np}\Gamma_{1,m+n}.
\end{eqnarray}
For the special case of $p=1$, we have
\begin{eqnarray}\label{e3.22}
\Delta_1&=&0\nonumber\\
\tau_1 &=&  \epsilon~(\frac{1}{g^2}~x +\frac{1}{3g}~x^3)\nonumber\\
\Delta_2&=& -\epsilon^2~\frac{11}{8g^4}\nonumber\\
\tau_2&=&\tau_1-\epsilon^2~(\frac{7}{8g^4}~x^2-\frac{1}{8g^3}~x^4).
\end{eqnarray}
The result of $p=0$ is already shown as the simple example at the
end of Section 2. These results are exactly the same as those
obtained based on the single trajectory quadrature method. After
expanding $e^{-\tau_n}$ to a power series the result is consistent
to the one based on the perturbation expansion using the same set
of Green function in ref.[2].

It is worthy to mention that the formula based on the Green
function $\overline{D}$ has another advantage for one-dimensional
case. Since the expression of $\tau_n$ is a double integration,
$\tau'_n$ could always be expressed as a single integration, i.e.,
no differentiation is needed. Comparing to the single trajectory
quadrature method introduced in ref[2], this revised iteration
method could reach the same result, however, it has much less
restriction to the potential. Because the existence of many times
of derivatives in the earlier method it requires the potential
being continuous and very smooth for high order derivatives, while
the revised iteration formula based on $\overline{D}$ is more
flexible and could even be applied to discontinuous potentials.\\

\noindent
{\bf 3.2 Stark Effect}\\

As a simple multi-dimensional example we now discuss the Stark
effect, making detailed comparison of the two iteration
procedures. Let $H$ be the unperturbed Hamiltonian for a Coulomb
potential:
\begin{eqnarray}\label{e3.23}
H = -\frac{1}{2} {\bf \nabla}^2 - \frac{g^2}{r},
\end{eqnarray}
with  ${\bf \nabla}^2 $ denoting the three-dimensional Laplacian, $r$
the radius. Consider the Stark effect as perturbation
the corresponding Hamiltonian is
\begin{eqnarray}\label{e3.24}
{\cal H} = H + \epsilon r {\rm cos}\vartheta,
\end{eqnarray}
where $\vartheta$ is the polar angle; i.e.,
\begin{eqnarray}\label{e3.25}
r^2 = x^2 + y^2 + z^2~~~~{\sf and}~~~~~z = r {\rm cos}\vartheta.
\end{eqnarray}
The ground-state of $H$, $\Phi(r)=e^{-S}=e^{-g^2r}$ satisfies
\begin{eqnarray}\label{e3.26}
H \Phi(r)= E \Phi(r) = -\frac{1}{2}g^4~\Phi(r)
\end{eqnarray}
and the ground-state wave function of ${\cal H}$ satisfies
\begin{eqnarray}\label{e3.27}
{\cal H} \Psi({\bf r}) = (E+\Delta) \Psi({\bf r}).
\end{eqnarray}
Introducing
\begin{eqnarray}\label{e3.28}
\Psi({\bf r}) = e^{-S(r)-\tau({\bf r})},
\end{eqnarray}
the perturbative energy $\Delta$ and the function $\tau({\bf r})$
introduced for the wave function in Eq.(3.28) could be solved
iteratively based on the revised iterative formula Eqs.(2.19) or
(2.32) for $\tau_n$ and Eq.(2.31) for $\Delta_n$. The
corresponding functions $C$ and $\overline{D}$ used in the Green
function $\overline{G}$ are expressed as (see Appendix C)
\begin{eqnarray}\label{e3.29}
C &=&\frac{1}{g^2} \int\limits_0^r~dr\\
\overline{D}&=&-2\int\limits_0^r~dr\frac{1}{r^2}~e^{2g^2r}
\int\limits_r^{\infty}~dr'r'^2 e^{-2g^2r'}.
\end{eqnarray}
Using the definitions of corresponding quantities in the spherical
coordinate given in Appendix C, we can perform the following two
iterations. Therevised one is
\begin{eqnarray}\label{e3.31}
\Delta_n&=&\frac{\int d{\bf r}~e^{-2g^2r} [~\epsilon r {\rm
cos}\vartheta - \frac{1}{2}~({\bf \nabla}\tau_{n-1})^2~]}{\int
d{\bf r}~e^{-2g^2r}}~,\\
\tau_n &=& (1+CT)^{-1}C~[~\epsilon r {\rm cos}\vartheta-\Delta_n-
\frac{1}{2}~({\bf \nabla}\tau_{n-1})^2~]\nonumber \\
       &=& (1+\overline{D}T_{\alpha})^{-1}\overline{D}~
       [~\epsilon r {\rm cos}\vartheta-\Delta_n-
\frac{1}{2}~({\bf \nabla}\tau_{n-1})^2~],
\end{eqnarray}
starting from $\Delta_0=0$ and $\tau_0({\bf r})=0$ and the old one
is
\begin{eqnarray}\label{e3.33}
\Delta_n&=&\frac{\int d{\bf r}~e^{-2g^2r} ~\epsilon r {\rm
cos}\vartheta f_{n-1}({\bf r})}{\int
d{\bf r}~e^{-2g^2r}f_{n-1}({\bf r})}~,\\
f_n &=& 1+(1+CT)^{-1}C~[\Delta_n-\epsilon r {\rm cos}\vartheta]f_{n-1}({\bf r})\nonumber \\
       &=&1+ (1+\overline{D}T_{\alpha})^{-1}\overline{D}~
       [\Delta_n-\epsilon r {\rm cos}\vartheta]f_{n-1}({\bf r}),
\end{eqnarray}
starting from $\Delta_0=0$ and $f_0({\bf r})=1$. From the revised
iteration formula (3.31) and (3.32) it is easy to obtain
\begin{eqnarray}\label{e3.35}
\Delta_1&=&\frac{\int d{\bf r}~e^{-2g^2r} [~\epsilon r {\rm
cos}\vartheta ]}{\int d{\bf r}~e^{-2g^2r}}=0~,\\
\tau_1 &=& \epsilon (\frac{1}{2g^2}r^2 +\frac{4}{(2g^2)^2}
       r ){\rm cos}\vartheta.
\end{eqnarray}
To calculate $\Delta_2$ and $\tau_2$ we need
\begin{eqnarray}\label{e3.37}
({\bf \nabla}\tau_1)^2 = \epsilon^2
 [(\frac{16}{(2g^2)^4} +\frac{8}{(2g^2)^3}~r+\frac{1}{(2g^2)^2}~r^2)
 +(\frac{8}{(2g^2)^3}~r+\frac{3}{(2g^2)^2}~r^2)
        {\rm cos}^2\vartheta]
\end{eqnarray}
and the obtained results are
\begin{eqnarray}\label{e3.38}
\Delta_2&=&\frac{\int d{\bf r}~e^{-2g^2r} [~\epsilon r {\rm
cos}\vartheta -\frac{1}{2}({\bf \nabla}\tau_1)^2
]}{\int d{\bf r}~e^{-2g^2r}}=-\epsilon^2 \frac{36}{(2g^2)^4}\\
\tau_2 &=& \epsilon (\frac{1}{2g^2}r^2 +\frac{4}{(2g^2)^2}
       r ){\rm cos}\vartheta\nonumber\\
       &&-\epsilon^2[\frac{1}{3(2g^2)^3}~r^3(1+3 {\rm
cos}^2\vartheta)
       +\frac{7}{(2g^2)^4}~r^2(1+ {\rm cos}^2\vartheta)].
\end{eqnarray}
Further iteration will give
\begin{eqnarray}\label{e3.40}
\Delta_4=\epsilon^2 \frac{36}{(2g^2)^4}+\epsilon^4 \frac{3555}{64g^{20}}
+O(\epsilon^6).
\end{eqnarray}
The result is exactly the same as those based on the single
trajectory quadrature method. The old iteration based on
Eqs.(3.33) and (3.34) gives
\begin{eqnarray}\label{e3.41}
\Delta_1&=&\frac{\int d{\bf r}~e^{-2g^2r} ~\epsilon r {\rm
cos}\vartheta}{\int d{\bf r}~e^{-2g^2r}}=0,\\
f_1({\bf r})&=&1-\epsilon(\frac{1}{sg^2}~r^2+\frac{4}{(2g^2)^2}~r)
{\rm cos}\vartheta.
\end{eqnarray}
Further iteration gives
\begin{eqnarray}\label{e3.43}
 \Delta_2&=&\frac{\int d{\bf r}~e^{-2g^2r}
~\epsilon r {\rm cos}\vartheta f_1({\bf r})}{\int d{\bf
r}~e^{-2g^2r}f_1({\bf r})}
         =-\epsilon^2\frac{36}{(2g^2)^4},\\
f_2({\bf
r})&=&1-\epsilon(\frac{1}{2g^2}~r^2+\frac{4}{(2g^2)^2}~r){\rm
       cos}\vartheta\nonumber\\
       &&+\epsilon^2[\frac{1}{3(2g^2)^3}~r^3(1+3 {\rm cos}^2\vartheta)
       +\frac{7}{(2g^2)^4}~r^2(1+ {\rm
       cos}^2\vartheta)]\nonumber\\
       &&+\frac{1}{2}\epsilon^2[(\frac{1}{2g^2}~r^2)^2
       +(\frac{4}{(2g^2)^2}~r)^2] {\rm
       cos}^2\vartheta+O(\epsilon^3).
\end{eqnarray}
Now we could see clearly, the perturbed wave function $e^{-\tau}$
in the revised iteration is an exponential while $f_n$ in the old
iteration is a polynomial, which could be meaningful only when
$\epsilon/g^2<<1$. It can be seen from Eqs.(3.39) and (3.44) that
$e^{-\tau_2}$ and $f_2$ are consistent to each other up to the
order of $\epsilon^2$, after making a power expansion for
$e^{-\tau_2}$. The energy correction for the 4th order from the
old iteration formula is expressed as a ratio of two polynomials.
However, after making the power series expansion for the factor
$\epsilon/g^2$, it gives also the same result as Eq.(3.40) given
from the revised formula, up to the order of $\epsilon^4$.
Therefore, to compare the results of the two iteration procedures
the power series expansion for the factor $\epsilon/g^2$ for both
cases are needed. Up to any fixed order of $\epsilon/g^2$ the two
results are consistent to each other, if the iteration reaches the
corresponding order. Of course, to keep the expansion meaningful
the condition $\epsilon/g^2<<1$ should be satisfied, although the
revised iteration itself does not need this condition, if one does
not make the expansion.

\newpage

\section*{\bf 3. Discussions}
\setcounter{section}{4} \setcounter{equation}{0}

In this paper a revised new iterative method based on Green
function defined by quadratures along a single trajectory is
proposed to solve the low-lying quantum wave function for
Schroedinger equation. The method is applied to solve the
unharmonic oscillator potential. A detailed comparison of the
revised iteration method to the old one is made using the example
of Stark effect.

From above analysis the advantages of this revised iteration
method can be seen clearly. Because it directly gives an
exponential form for the perturbed wave function $e^{-\tau}$, it
sometimes gives better convergence than the old formula, like the
case of unharmonic oscillator shown at the end of Section 2. The
exponential form of the perturbed wave function also ensures the
result of this revised iteration procedure consistent with those
obtained using the series expansion of $\{S_i\}$ and $ \{E_i\}$
based on the single trajectory quadrature(See Section 1 of
Ref.[2]). When we look at the formula Eq.(2.30) for the
calculation of the perturbation energy, it is completely new and
much simpler than the one in the old iteration procedure (see
Eq.(2.17)). The expression of the perturbation energy in the old
iteration method is similar to the one in the usual perturbation
method. In the new iteration formula Eq.(2.30) the denominator
does not change in each order of iteration. This makes the
calculation much simpler. The improvement for the energy and the
wave function in each order of iteration is very clear. This
revised iteration procedure will be applied to more other
potentials in our future work.

\section*{\bf Acknowledgment}

The author is grateful to Professor T. D. Lee for his continuous
and substantial instructions and advice. This work is partly
supported by National Natural Science Foundation of China (NNSFC,
No. 20047001).

\section*{\bf Reference}

1. R. Friedberg, T. D. Lee and W. Q. Zhao, IL Nuovo Comento
A112(1999)1195

\noindent 2. R. Friedberg, T. D. Lee and W. Q. Zhao, Ann. Phys.
288(2001)52

\noindent
3. R. Friedberg, T. D. Lee, W. Q. Zhao and A. Cimenser, Ann. Phys. 294(2001)67\\

\newpage

\noindent
{\bf Appendix A}\\

For the convenience of applying the coordinate system
$\{S,~\alpha\}$ the definition introduced in ref.[2] is given in
the following. Based on the set of $N-1$ angular variables defined
in Eq.(2.11)
$$
\alpha = (\alpha_1({\bf q}),\alpha_2({\bf q}), \cdots, \alpha_{N-1}({\bf q})),
\eqno(A.1)
$$
each point ${\bf q}$ in the $N$-dimensional space will now be designated by
$$
(S, \alpha_1, \alpha_2, \cdots, \alpha_{N-1}),
\eqno(A.2)
$$
instead of $(q_1, q_2, q_3, \cdots, q_N)$.
The corresponding line element can be written as
$$
d\stackrel{\rightarrow}{{\bf q}} =
\stackrel{\wedge}S h_S dS
+\sum_{j=1}^{N-1}\stackrel{\wedge}{\alpha}_j h_j d\alpha_j;
\eqno(A.3)
$$
the gradient is given by
$$
{\bf \nabla} = \stackrel{\wedge}S
\frac{1}{h_S}\frac{\partial}{\partial S}
+\sum_{j=1}^{N-1}\stackrel{\wedge}{\alpha}_j \frac{1}{h_j}
\frac{\partial}{\partial \alpha_j},
\eqno(A.4)
$$
and
$$
T=-\frac{1}{2} {\bf \nabla}^2
\eqno(A.5)
$$
can be decomposed into two parts:
$$
T = T_S + T_{\alpha},
\eqno(A.6)
$$
with
$$
T_S = -\frac{1}{2h_Sh_\alpha} \frac{\partial}{\partial S}
(\frac{h_{\alpha}}{h_S}\frac{\partial}{\partial S}),
\eqno(A.7)
$$
$$
T_{\alpha} = -\frac{1}{2h_Sh_\alpha} \sum_{j=1}^{N-1}
\frac{\partial}{\partial \alpha_j}
(\frac{h_Sh_{\alpha}}{h_j^2}\frac{\partial}{\partial \alpha_j}),
\eqno(A.8)
$$
in which
$$
h_{\alpha} = \prod_{j=1}^{N-1}h_j,
\eqno(A.9)
$$
and
$$
h_S^2 = [({\bf \nabla}S)^2]^{-1},~~~
h_1^2 = [({\bf \nabla}\alpha_1)^2]^{-1},\cdots,~~~
h_j^2 = [({\bf \nabla}\alpha_j)^2]^{-1},\cdots.
\eqno(A.10)
$$
The  volume element in the ${\bf q}$-space is
$$
d^N {\bf q} = h_Sh_\alpha dSd\alpha
\eqno(A.11)
$$
with
$$
d\alpha = \prod_{j=1}^{N-1}d\alpha_j.
\eqno(A.12)
$$

\noindent
{\bf Appendix B}\\

In the calculation of $\Delta_n$ and $\tau_n$ in Section 3 the
following expressions introduced in ref.[2] are needed. With the
definition of $C$ and $\overline{D}$ introduced in Eqs.(2.39) and
(2.47), for $n>0$,
$$
Cx^{2n} = \int_0^S \frac{dS}{g2S}x^{2n} = \frac{1}{g(2n)} x^{2n}.
\eqno(B.1)
$$
$$
\overline{D}[x^{2n} - \Gamma_{1~n}]=(1+CT)^{-1}C[x^{2n} -
\Gamma_{1~n}] = \sum_{m=1}^n \Gamma_{m~n}x^{2m}.
\eqno(B.2)
$$
where
$$
\Gamma_{n~n} = \frac{1}{g2n},~~~\Gamma_{n-1~n} = \frac{2n-1}{2g^2(2n-2)},
$$
$$
\Gamma_{m~n} = \frac{(2n-1)(2n-3)\cdots(2m+1)}{m(2g)^{n-m+1}}~~~~
{\sf for}~~~~1 \leq m \leq n-1
\eqno(B.3)
$$
and in particular
$$
\Gamma_{1~n} = \frac{(2n-1)!!}{(2g)^n}.
$$
For $n \geq 0$,
$$
Cx^{2n+1} = \int_0^S \frac{dS}{g2S}x^{2n+1} = \frac{1}{g(2n+1)} x^{2n+1}.
\eqno(B.4)
$$
It is straightforward to establish the following:
$$\overline{D}x^{2n+1}=(1+CT)^{-1}Cx^{2n+1} = \sum_{m=0}^n \gamma_{m~n}x^{2m+1}
\eqno(B.5)
$$
where
$$
\gamma_{n~n} = \frac{1}{g(2n+1)},~~~\gamma_{n-1~n} = \frac{n}{g^2(2n-1)},
$$
$$
\gamma_{m~n} = \frac{n(n-1)\cdots(m+1)}{g^{n-m+1}(2m+1)}~~~~
{\sf for}~~~~0 \leq m < n
\eqno(B.6)
$$
and in particular
$$
\gamma_{0~n} = \frac{n!}{g^{n+1}}.
$$

It is convenient to extend the definitions of $\gamma_{m~n}$
and $\Gamma_{m~n}$, by defining
$$
\gamma_{m~n}=\Gamma_{m~n}=0~~~~~~{\sf~for}~~~~m>n
\eqno(B.7)
$$
and
$$
\Gamma_{0~n}=0.
$$

\newpage

\noindent
{\bf Appendix C}\\

For the special case of spherical coordinate in three-dimension
and for the coulomb potential we have $S=g^2r$. Correspondingly,
$$
\alpha=(\vartheta,~\varphi)
\eqno(C.1)
$$
$$
h_S=1,~~~h_{\vartheta}=r,~~~h_{\varphi}=r {\rm sin}\vartheta
\eqno(C.2)
$$
For any function $f({\bf r})$ we have
$$
({\bf \nabla}f({\bf r}))^2=(\frac{\partial f}{\partial r})^2+
(\frac{1}{r}\frac{\partial f}{\partial \vartheta})^2
+(\frac{1}{r {\rm sin}\vartheta}\frac{\partial f}{\partial \varphi})^2.
\eqno(C.3)
$$
$$
T=T_S+T_{\alpha}
\eqno(C.4)
$$
with
$$
T_S=-\frac{1}{2}\frac{1}{r^2}\frac{d}{dr}r^2\frac{d}{dr},
$$
$$
T_{\alpha}=\frac{1}{2r^2}L^2
\eqno(C.5)
$$
and $L^2$ is the operator of the square of the angular momentum.
The Green functions used in the iteration are defined as
$$
\overline{G}=(1+
\overline{D}T_{\alpha})^{-1}\overline{D}=(1+CT)^{-1}C.
\eqno(C.6)
$$
Assuming $\Psi(0)=\Phi(0)$ and $\Psi(r\rightarrow \infty)=0$, we have
$$
C =\frac{1}{g^2} \int\limits_0^r~dr
\eqno(C.7)
$$
$$
\overline{D}=-2\int\limits_0^r~dr\frac{1}{r^2}~e^{2g^2r}
\int\limits_r^{\infty}~dr'r'^2 e^{-2g^2r'}.
\eqno(C.8)
$$
In the iteration the following formula are applied:
$$
(1+CT)^{-1}Cr^n=\frac{1}{g^2}\sum_{m=2}^{n+1} \Lambda_{m,n}r^m
+\Lambda_{1,n}(1+CT)^{-1}C\cdot 1,
\eqno(C.9)
$$
$$
\overline{D}r^n=\frac{1}{g^2}\sum_{m=2}^{n+1} \Lambda_{m,n}r^m
+\Lambda_{1,n}\overline{D}\cdot 1,
\eqno(C.10)
$$
where
$$
\Lambda_{n+1,n}=\frac{1}{n+1}
$$
$$
\Lambda_{m,n}=\frac{1}{(2g^2)^{n+1-m}}~\frac{(n+2)(n+1)\cdots(m+2)}{m}
$$
$$
\Lambda_{1,n}=\frac{1}{(2g^2)^n}(n+2)(n+1)\cdots 3.
\eqno(C.11)
$$
Although $\overline{D}\cdot 1=(1+CT)^{-1}C\cdot 1$ is divergent, all terms
including this factor are cancelled in the iteration.
Defining ${\rm cos}\vartheta=\xi$ we have
$$
r^2 T_\alpha \xi=\xi
$$
and
$$
r^2 T_\alpha \xi^m = m(\frac{m+1}{2} \xi^m -\frac{m-1}{2} \xi^{m-2}).
\eqno(C.12)
$$
For Legendre function $P_l(\xi)$, we have
$$
r^2T_\alpha P_l(\xi)=\frac{l(l+1)}{2}~P_l(\xi).
\eqno(C.13)
$$
Based on the above formula the two iteration procedures could be performed.

\end{document}